\documentclass[aps,twocolumn,superscriptaddress,showpacs,groupedaddress,preprintnumbers]{revtex4}
\usepackage{amsfonts,amscd,amsmath, amssymb,graphicx,color}
\def\ep{\text{e}}
\def\g{\mathfrak{g}}
\def\oh{\frac{1}{2}}
\def\s{\mathfrak{s}}
\def\Fd{F_\text{drag}}
\def\rh{r_h}
\def\rv{r_v}
\def\wv{w_v}
\def\wh{w_h}
\begin{document}
\preprint{LMU-ASC 40/17}
\title{Drag Force on Heavy Quarks and Spatial String Tension}
\author{Oleg Andreev}
 \affiliation{L.D. Landau Institute for Theoretical Physics, Kosygina 2, 119334 Moscow, Russia}
\affiliation{Arnold Sommerfeld Center for Theoretical Physics, LMU-M\"unchen, Theresienstrasse 37, 80333 M\"unchen, Germany}
\begin{abstract} 
Heavy quark transport coefficients in a strongly coupled Quark-Gluon Plasma can be evaluated using a gauge/string duality and lattice QCD. Via this duality, one can argue that for low momenta the drag coefficient for heavy quarks is proportional to the spatial string tension. Such a tension is well studied on the lattice that allows one to straightforwardly make non-perturbative estimates of the heavy quark diffusion coefficients near the critical point. The obtained results are consistent with those in the literature.
\end{abstract}
\maketitle

Heavy quarks are one of the most valuable probes to study the properties of a strongly coupled Quark-Gluon Plasma (sQGP) in heavy ion collision experiments \cite{Q-review}. When a heavy quark moves through the plasma, it feels a drag force and consequently loses energy. The magic of AdS/CFT is that one can use the classical picture of a trailing string in a five-dimensional spacetime to capture strong coupling dynamics of this process \cite{drag,book}. While it is a good starting point, one must keep in mind that replacing QCD by $N = 4$ super-Yang-Mills is not appropriate and might be completely misleading. Sadly, there still is no string theory which provides a dual description of QCD and that is why in practice one is led to consider effective string theory on deformed AdS spacetimes to somehow model QCD.

A long standing question is: In what ways can string theory be useful in understanding the physics of heavy ion collisions? In this note we propose a new way to make non-perturbative estimates of the heavy quark transport coefficients near the critical point as one more step toward answering this question. 

The approach will follow the strategy of describing QCD by means of five-dimensional effective string theory. In doing so, we consider a Nambu-Goto string whose action is $S=-\tfrac{1}{2\pi\alpha'}\int d\tau d\sigma\sqrt{-\gamma}$, on a five-dimensional spacetime with a metric

\begin{equation}\label{metric}
ds^2=w(r) R^2\Bigl(-f(r)dt^2+dx_i^2+f^{-1}(r)dr^2\Bigr)
\,.
\end{equation}
Here $x_i=(x,y,z)$. We think of this spacetime as a deformation of the Schwarzschild black hole in $\text{AdS}_5$ space such that the boundary is at $r=0$ and the horizon at $r=\rh$. As $r$ approaches the boundary, the metric approaches that of the Schwarzschild black hole with $w=\frac{1}{r^2}$ and $f=1-\bigl(\tfrac{r}{\rh}\bigr)^4$. In this limit, $R$ becomes the anti-de Sitter radius. We assume that this string model provides a reasonable approximation to the behavior of QCD in the deconfined phase near the critical (crossover) temperature at zero baryon chemical potential \cite{ads/qcd}. As usual, the Hawking temperature of the black hole is identified with 
the temperature of the gauge theory dual such that $T=\frac{1}{4\pi}\lvert \frac{d f}{d r}\rvert_{r=\rh}$. We also assume that the plasma is isotropic, and because of this the metric is chosen to be invariant under spatial rotations. 

\begin{figure*}[htbp]
\centering
\includegraphics[width=5.15cm]{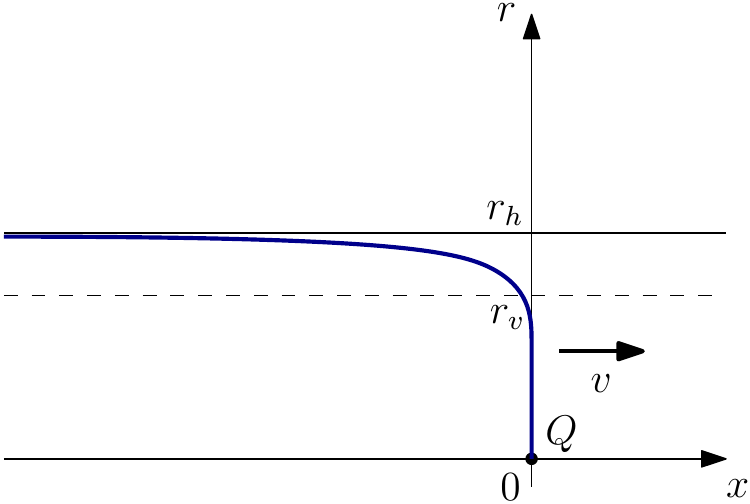}
\hspace{3cm}
\includegraphics[width=5.25cm]{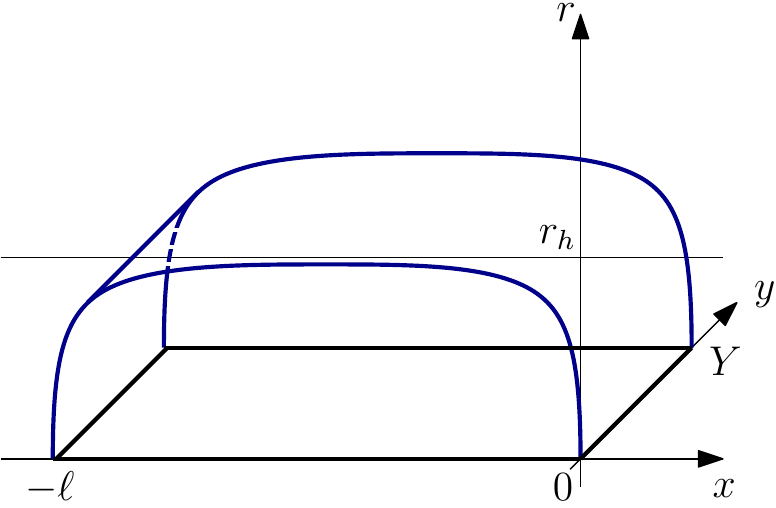}
\caption{{\small Left: A string trailing out from the heavy quark $Q$ into the black hole horizon. 
Right: A minimal surface for a rectangular Wilson loop of length $\ell$ and width $Y$. The loop lies on the $xy$-plane at $r=0$. For large $Y$, a dominant contribution comes from a lateral surface whose area is proportional to $Y$.}}
\label{trailing}
\end{figure*}

Now let us discuss a string attached to an external quark that moves with speed $v$ in the $x$ direction. We assume that the quark bare mass is very large so that the quark is on the boundary at $r=0$, as shown in Figure \ref{trailing} on the left. Then the drag coefficient is calculated from the momentum flow flowing from the boundary to the horizon along the string worldsheet, as originally described in \cite{drag} for AdS space. The arguments in the generic (non-conformal) case \eqref{metric} are similar to those in the AdS case except that the effective string tension \cite{az3}, $\sigma_{\text{eff}}=w\sqrt{f}$, must be a decreasing function of $r$ on the interval $[0,\rh]$ to have a single solution. If so, then the drag force is given by \cite{drag-kir0} 

\begin{equation}\label{drag-d}
\Fd=-\g\wv v
\,,
\end{equation}
where $\g=\frac{R^2}{2\pi\alpha'}$, $\wv=w(\rv )$, and $\rv$ is a solution of the equation $f(\rv )=v^2$.

What is puzzling about this trailing string picture is that the string, despite being infinitely long, does not break in the deconfined phase. How can it be possible to avoid breaking? Before answering this question, let us make a detour and discuss the spatial Wilson loops. The properties of those were well studied in lattice QCD \cite{lattice-review}. The main point is that the spatial Wilson loops always, even above $T_c$, obey an area law. For the simple case of a rectangular loop, shown in Figure \ref{trailing} on the right, it means that the asymptotic behavior is given by  

\begin{equation}\label{wilson-loop}
\langle\,W(\ell,Y)\,\rangle \sim\ep^{-\sigma_s\ell Y}
\,,\quad\text{as}\quad Y,\ell\rightarrow\infty
\,.
\end{equation}
Here $\sigma_s$ is the spatial string tension. In the context of AdS/CFT, the spacial Wilson loops were discussed in \cite{witten}. The computation is in fact reduced to finding minimal area surfaces in AdS space. The large distance physics is determined by the near horizon geometry of the black hole so that the spatial string tension is written as a function of $\rh$. This is illustrated in Figure \ref{trailing}, on the right. By essentially the same arguments, it can be shown that this is the case whenever the factor $w$ in \eqref{metric} is a decreasing function of $r$ on the interval $[0,\rh]$ (see, e.g., \cite{sst}). The tension now takes the form

\begin{equation}\label{ss}
\sigma_s=\g\wh 
\,,
\end{equation}
with $\wh=w(\rh )$.

\begin{figure*}[htbp]
\centering
\includegraphics[width=7cm]{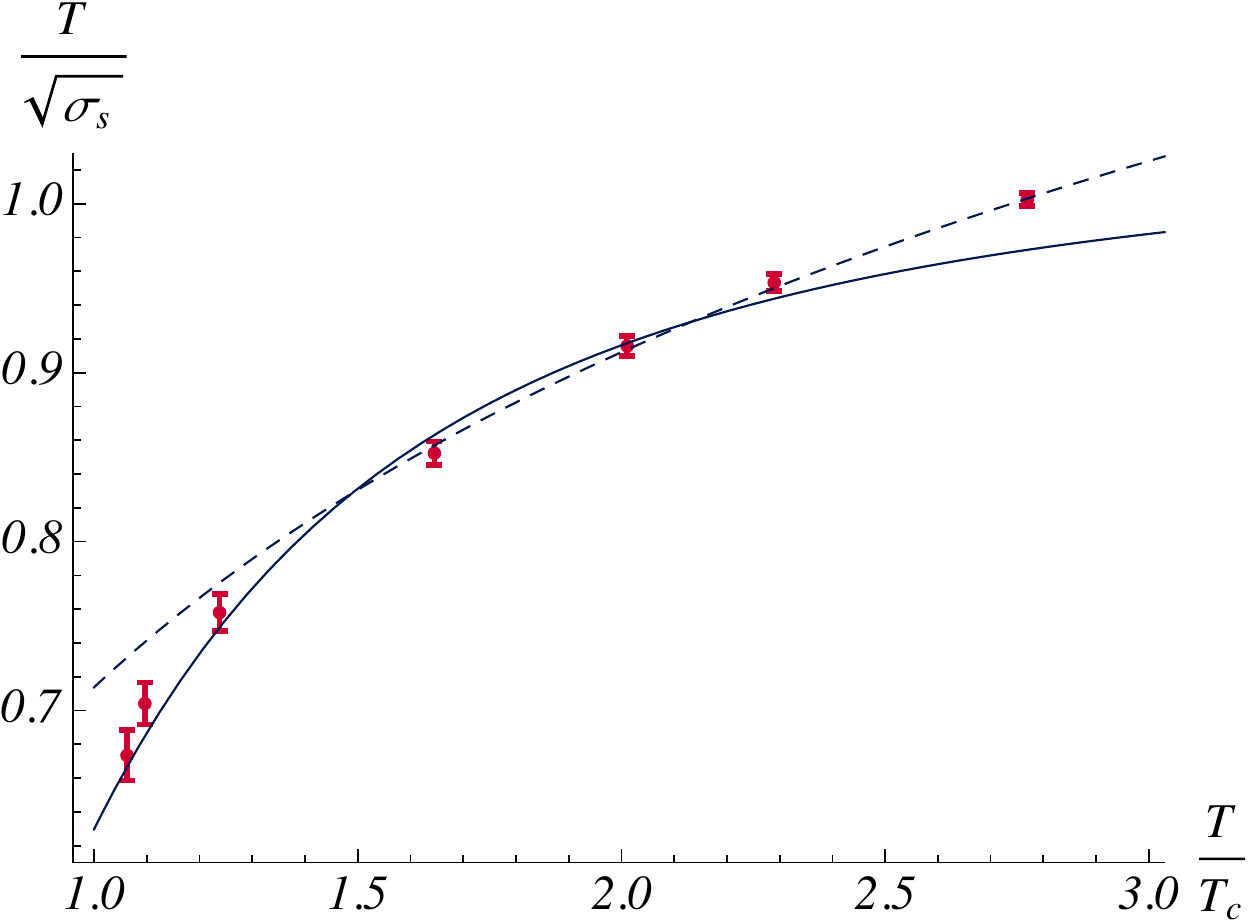}
\hspace{2.5cm}
\includegraphics[width=7cm]{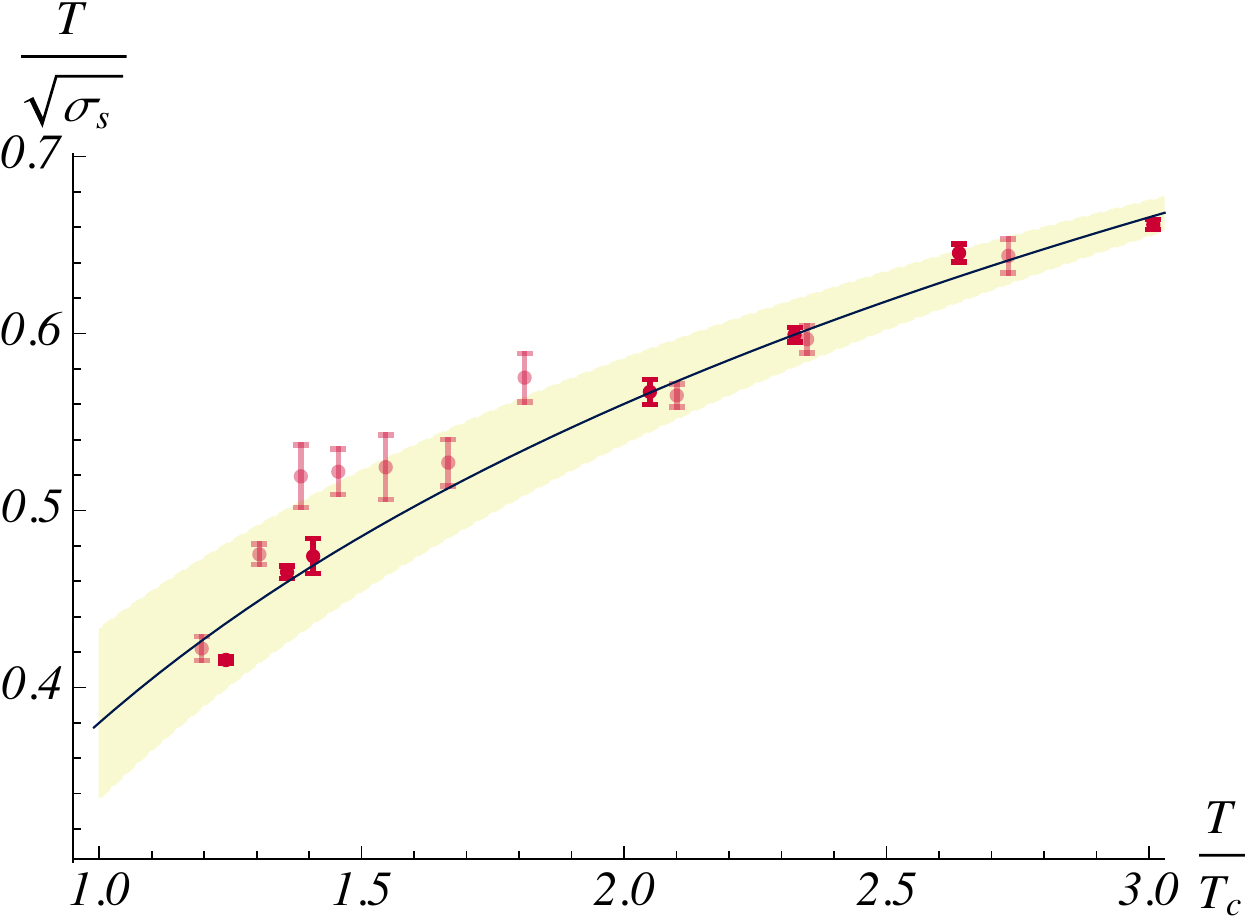}
\caption{{\small The temperature over the square root of the spatial string tension versus $\tfrac{T}{T_c}$ for $SU(3)$. Left: Quenched QCD. The dots are from lattice simulations of \cite{lattice0}. The solid curve corresponds \eqref{sst0ads} with $\g=0.094$ and the dashed curve to \eqref{sst0} with $c=0.566$ and $\Lambda=0.104\,T_c$. Right: $(2+1)$-flavor QCD. The value of $T_c$ is fixed to be $155\,\text{MeV}$. The dots are calculated on $N_\tau=6$ (light) and $N_\tau=8$ (dark) lattices \cite{lattice2+1}. The solid curve corresponds to \eqref{sst2+1} with $a_0=0.38$ and $a_1=0.26$. The yellow band is used to indicate uncertainties. It includes the numerical values calculated on both lattices.}}
\label{sst}
\end{figure*}

So far, our analysis has been carried out without any reference to the number of colors, but it is important for what follows to know the explicit form of $\sigma_s$ as a function of $T$ for three colors. We begin by briefly reviewing two parameterizations of the lattice data for pure $SU(3)$ gauge theory (quenched QCD) \cite{lattice0}. The parameterization of \cite{sst}

\begin{equation}\label{sst0ads}
\frac{T}{\sqrt{\sigma_s}}=\frac{1}{\pi\sqrt{\g}}\exp\biggl(-\oh\frac{T^2_c}{T^2}\biggr)
\,
\end{equation}
follows from \eqref{ss} with $w=\frac{\ep^{\s r^2}}{r^2}$ and $f=1-\bigl(\tfrac{r}{\rh}\bigr)^4$. It has one free parameter, $\g$. The critical temperature is determined as a function of a deformation parameter $\s$, or more explicitly $T_c=\frac{\sqrt{\s}}{\pi}$. The second parameterization, motivated by the renormalization group equation at two loops \cite{lattice0}, is given by 

\begin{equation}\label{sst0}
\frac{T}{\sqrt{\sigma_s}}=\frac{1}{c}\biggl(2b_0\ln\frac{T}{\Lambda}+\frac{b_1}{b_0}\ln\Bigl(2\ln\frac{T}{\Lambda}\Bigr)\biggr)
\,,
\end{equation}
with $b_0=\frac{11}{(4\pi)^2}$ and $b_1=\frac{102}{(4\pi)^4}$. It has two free parameters, $c$ and $\Lambda$. In the left panel of Figure \ref{sst}, we compare both parameterizations with the lattice. 
Obviously, string theory provides the better description near the phase transition point, but becomes worse for higher temperatures where the temperature dependence of the spatial string tension is determined by the renormalization group $\beta$-function of gauge theory. It is natural to think of these parameterizations as two complementary descriptions: one for the strong coupling regime and another for weak. Of course, what is of primary importance is the properties of $\sigma_s$ in $(2+1)$-flavor QCD. For this case \cite{lattice2+1} there also exists the parameterization motivated by the renormalization group equation at two loops but even simpler than that is 

\begin{equation}\label{sst2+1}
\frac{T}{\sqrt{\sigma_s}}=a_0+a_1\ln\frac{T}{T_c}
\,,
\end{equation}
where $T_c$ is the crossover temperature. It also has two free parameters, $a_0$ and $a_1$. The essential point to be noticed is that the results obtained for $T_c$ are still not very satisfactory. Different observables lead to different numerical outcomes \cite{Tc}. In the case of the physical strange quark mass and almost physical light quark masses, the temperature dependence of the 
spatial string tension is shown in the right panel of Figure \ref{sst}. We see that the simple parameterization is not bad at all and might be quite useful for practical purposes. However, neither here nor in string theory is it well understood what is going on with the crossover. We believe that these issues are worthy of future study \cite{ads/qcd}.

Now going back to an arbitrary number of colors, we can draw an interesting deduction about the form of the drag force on heavy quarks in the non-relativistic limit (low momenta). To leading order, from the expression \eqref{drag-d} we have 

\begin{equation}\label{drag-dn}
\Fd=-\g\wh v+O(v^3)\,,
\end{equation}
where we have used the fact that if $v\rightarrow 0$, then $\rv=\rh -\frac{1}{4\pi}\frac{v^2}{T}+O(v^3)$ is a solution of the equation $f(\rv)=v^2$. In terms of $\sigma_s$, that means 

\begin{equation}\label{drag-ds}
\Fd=-\sigma_s v+O(v^3)\,,
\end{equation}
as follows from equation \eqref{ss}. Then the drag coefficient for heavy quarks is given by

\begin{equation}\label{eta}
\eta_{\scriptscriptstyle D}=\frac{\sigma_s}{M}
\,.
\end{equation}
Here $M$ is a heavy quark "kinetic mass". This is our key result which extends that of \cite{sin} for $w=\frac{1}{r^2}$ to the generic case of $w(r)$, but for low momenta. On the string theory side it is background independent (universal) in the sense that the result holds for the class of five-dimensional geometries such that $w\sqrt{f}$ and $w$ are decreasing functions of $r$ on the interval $[0,\rh]$. On the gauge theory side it assumes only that the string models on these geometries provide a reasonable approximation to QCD in the deconfined phase near $T_c$ at zero baryon chemical potential. 

It is instructive to analyze the behavior of the trailing solution in the presence of an external force. This force will stretch the string a distance $\Delta x=v\Delta t$ over a short time interval $\Delta t$. In the non-relativistic limit, the energy gain of the string is calculated from the near horizon geometry alone and given by $\Delta E=\g\wh v\Delta x$. Then, with the help of equation \eqref{ss}, one can show that the effective tension for a string stretched along the horizon is simply $\sigma_\text{eff}=\sigma_s v$. This tension is always non-zero. Therefore, no string breaking occurs and the trailing solution makes sense for theories above $T_c$, with or without dynamical quarks. 

Since we do not have a satisfactory framework in which to describe $M$, we convert the result into the diffusion coefficients via the Einstein relations \cite{Q-review}. So, for low momenta we have 

\begin{equation}\label{Ds}
D_s=\frac{T}{\sigma_s}
\,,
\end{equation}
with $D_s$ the spatial diffusion coefficient, and 

\begin{equation}\label{Dm}
D=T\sigma_s
\,,
\end{equation}
with $D$ the momentum diffusion coefficient. The latter is simply related to the momentum broadening coefficient $\kappa$ by $D=\frac{\kappa}{2}$ \cite{book}.

\begin{figure*}[htbp]
\centering
\includegraphics[width=7cm]{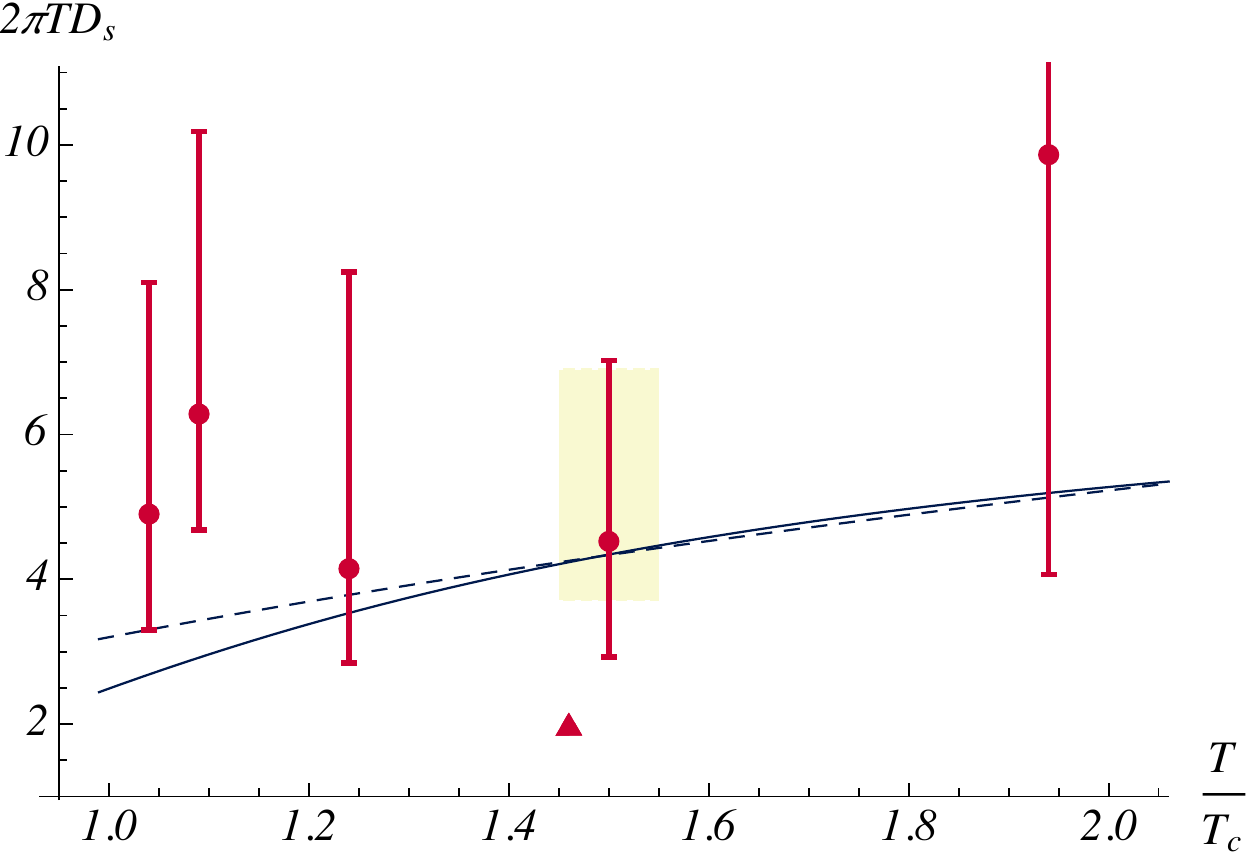}
\hspace{2.5cm}
\includegraphics[width=7cm]{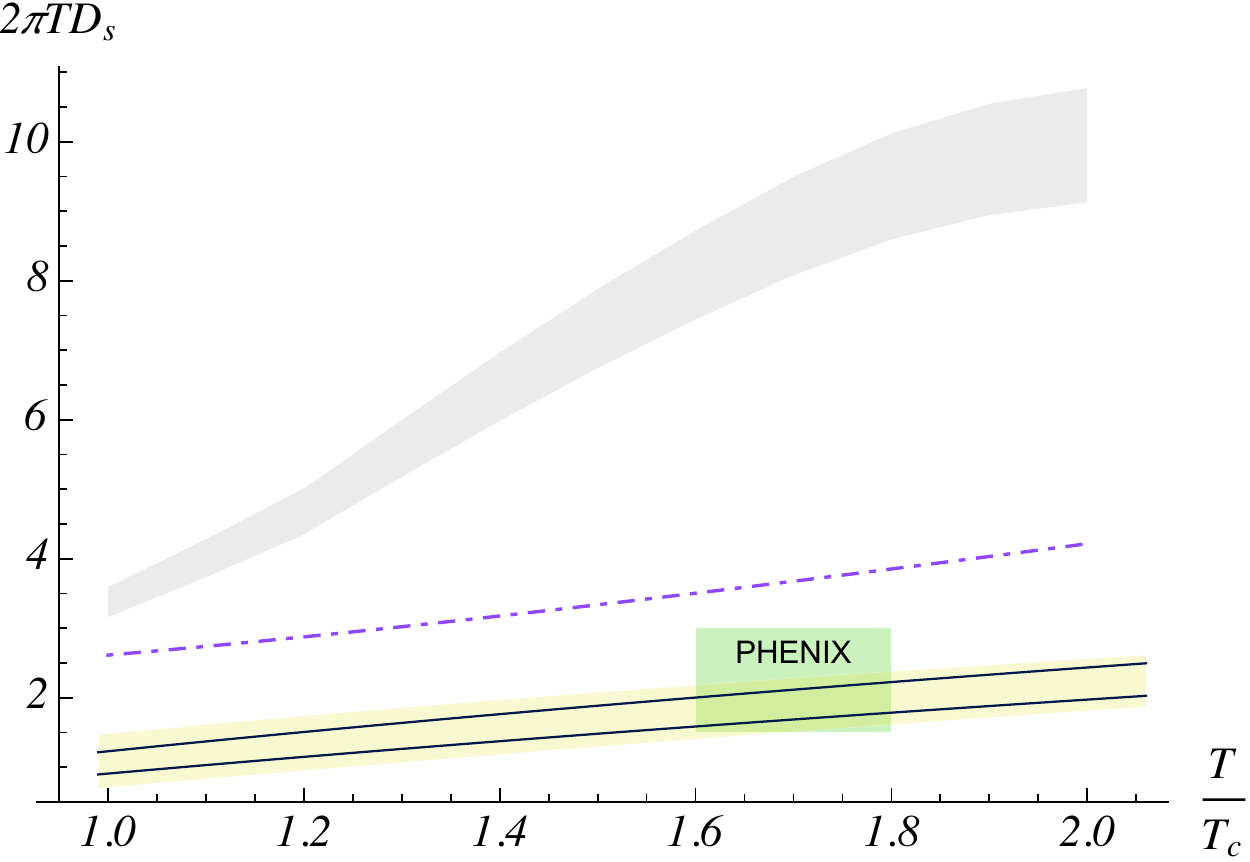}
\caption{{\small The spatial diffusion coefficient, scaled with the thermal wavelength, versus $\frac{T}{T_c}$ for $SU(3)$. Left: Quenched QCD. The dots are from lattice simulations of \cite{datta} and the triangle from \cite{kar}. The yellow range illustrates the estimate of \cite{kac}. The solid and dashed curves represent our estimates based on the formula \eqref{Ds} and the parameterizations of Figure \ref{sst}. Right: Full QCD. The upper band and dash-dotted line are resulting from the $T$-matrix \cite{rapp} and perturbative QCD calculations with the reduced Debye mass and running coupling \cite{nantes}. The range preferred by the $v_2$ measured by PHENIX is also shown, as in Figure 6 of \cite{datta}. The solid curves represent our estimates obtained from the formula \eqref{Ds} and parameterization of Figure \ref{sst} at $T_c=155\,\text{MeV}$ (bottom) and $T_c=196\,\text{MeV}$ (top). Like in Figure \ref{sst}, the yellow band is used to indicate uncertainties.}}
\label{diffs}
\end{figure*}

\begin{figure}[htbp]
\centering
\includegraphics[width=7cm]{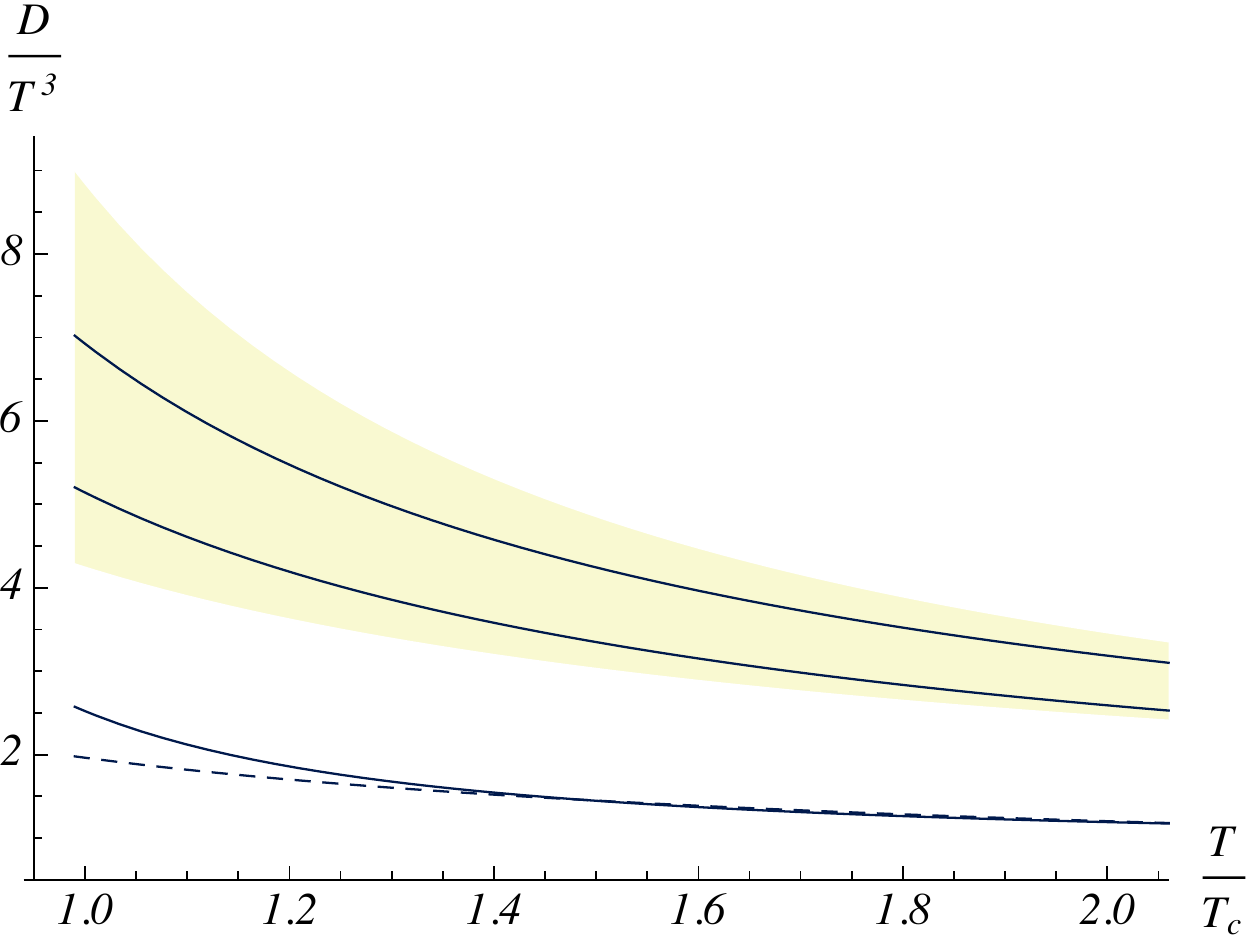}
\caption{{\small The momentum diffusion coefficient over the cube of temperature versus $\frac{T}{T_c}$ for $SU(3)$. The two bottom curves represent the estimates for quenched QCD, while the two others for $(2+1)$-flavor QCD at $T_c=155\,\text{MeV}$ (top) and $T_c=196\,\text{MeV}$ (bottom). As before, the yellow band is used to indicate uncertainties.}}
\label{D}
\end{figure}

It is of great interest to compare our predictions for the temperature dependence of the diffusion coefficients with other results in the literature. There is an obvious question to be asked at this point. How exact are the formulas \eqref{Ds} and \eqref{Dm}? No real answer will be proposed here. All that we can do is to ask whether those are reasonable from a quantitative point of view. Since string theory usually provides a good approximation in the deconfined phase near the critical point for $T_c\lesssim T\lesssim 3T_c$ \cite{ads/qcd,sst}, we will restrict out attention to this interval. We begin with the spatial diffusion coefficient. It seems an interesting fact that scaling $D_s$ with the thermal wavelength of medium $\lambda = \frac{1}{2\pi T}$ leads to a dimensionless quantity which is related to the quantity studied in lattice QCD by $2\pi TD_s=2\pi\Bigl(\frac{T}{\sqrt{\sigma_s}}\Bigr)^2$. And so, let us first consider the case of quenched lattice QCD. A summary of the results is shown in Figure \ref{diffs} on the left. 
Both our estimates are substantially of the same order of magnitude as those numerically calculated from two-point correlation functions. However, the lattice results of \cite{datta,kar,kac} are currently inconclusive, so it is difficult to draw a definite conclusion on how accurate or inaccurate our predictions are. The situation becomes even more involved if dynamical quarks are present. In this case, the dimensionless quantity is estimated to be in a wide range of $2\pi TD_s\simeq 1.5-32$ \cite{Q-review,datta}. The right panel of Figure \ref{diffs} shows some of the estimates close to the lower bound of this range. Ours being on its boundary are in a 
quite good agreement with the estimate of \cite{datta} based on the data measured by PHENIX at RHIC \cite{phenix}. As seen, the effect of dynamical quarks is two-fold: the obtained values of $D_s$ are smaller than those for quenched QCD, and in addition there is a slight decrease in growth with temperature. It is worth noting that like in AdS/CFT the quantity $TD_s$ is the same for $c$ and $b$ quarks, but unlike that it is temperature dependent, as expected in QCD. Having discussed the spatial diffusion coefficient, it is straightforward to see what happens to the momentum diffusion coefficient with the help of a simple inversion formula $\frac{D}{T^3}=(TD_s)^{-1}$. In this note we will not dig deep enough into this topic, but for completeness, present our estimates in Figure \ref{D}. 
All these are based on the formula \eqref{Dm} and the parameterizations of Figure \ref{sst}. 

In conclusion, we have found that near the critical point the drag force coefficient for heavy quarks at low momentum is simply related to the spatial string tension, which so far was a classic non-perturbative probe for the convergence of the weak coupling expansion at high temperatures. Now, new and promising opportunities arise by extracting the diffusion coefficients from lattice simulations or from dimensionally reduced QCD \cite{laine}. We believe that our results provide some clue to answering the question posed at the beginning of this note, but still not enough.

This work was supported in part by Russian Science Foundation Grant 
No.16-12-10151. We are grateful to S. Hofmann, I. Sachs, P. Weisz, and U.A. Wiedemann for helpful discussions, and M. He, P. Petreczky, and R. Rapp for providing us with the numerical data used in Figures \ref{sst} and \ref{diffs}. We also would like to thank the Arnold Sommerfeld Center for Theoretical Physics for hospitality during different stages of this work.


\end{document}